# SRDCNN: Strongly Regularized Deep Convolution Neural Network Architecture for Time-series Sensor Signal Classification Tasks

Arijit Ukil
TCS Research and Innovation
Tata Consultancy Services
Kolkata, India
arijit.ukil@tcs.com.com

Antonio Jara
University of Applied Sciences
Western Switzerland (HES-SO)
Switzerland
jara@ieee.org

Leandro Marin
University of Murcia Spain
leandro@um.es

## ABSTRACT

Deep Neural Networks (DNN) have been successfully used to perform classification and regression tasks, particularly in computer vision based applications. Recently, owing to the widespread deployment of Internet of Things (IoT), we identify that the classification tasks for time series data, specifically from different sensors are of utmost importance. In this paper, we present SRDCNN: Strongly Regularized Deep Convolution Neural Network (DCNN) based deep architecture to perform time series classification tasks. The novelty of the proposed approach is that the network weights are regularized by both $L^1$ and $L^2$ norm penalties. Both of the regularization approaches jointly address the practical issues of smaller number of training instances, requirement of quicker training process, avoiding overfitting problem by incorporating sparsification of weight vectors as well as through controlling of weight values. We compare the proposed method (SRDCNN) with relevant state-of-the-art algorithms including different DNNs using publicly available time series classification benchmark (the UCR/UEA archive) time series datasets and demonstrate that the proposed method provides superior performance. We feel that SRDCNN warrants better generalization capability to the deep architecture by profoundly controlling the network parameters to combat the training instance insufficiency problem of real-life time series sensor signals.

## 1. INTRODUCTION

We are witnessing a constant rise of demand for spectrum of applications like remote diagnosis for disease screening, automated detection machine fault, gesture recognition for assisting robot movement, activity detection for elderly care, and many more that can potentially improve the human quality of life. With the increasing availability of sensors that capture different useful physical signals, cyber and physical worlds are made closely coupled. Further, Internet of Things (IoT) provides the necessary infrastructure to enable the deployment of gamut of human-centric applications. One of the prime requirements of large number of human-centric applications is to perform the classification tasks. For example, 1. To detect whether the subject has cardiac disease like atrial fibrillation from Electrocardiogram signals, 2. To detect faulty automobile engine from its noise signal, 3. water quality detection from water distribution piping systems from unobtrusive examination of chlorine concentration etc.

We observe that majority of such tasks are classification of events from time series sensor signals. It is to be kept in mind that practical applications and development efforts to perform analytics operation on time series sensor signals require considerations of number of constraints like smaller number of training instances, faster learning on training datasets, incorporating the generalizability of the algorithm by avoiding the overfitting problem. In this paper, we propose a robust classification approach to analyze such time series signals under the given practical constraints. For example, in Table 1, we illustrate few of the available time series datasets for training purposes [1][2]. It is witnessed that practical time series sensor datasets pose significant challenge during training or learning task. We find that: 1. The total number of training instances is mostly less than 1000 considering altogether the training datasets from all the classes. Essentially, in many scenarios, typical number of training instances in each of the classes is less than 100. 2. The number of samples at each of the training instances are also too small, mostly below 500. The exemplary cases as perceived from Table 1, it is noteworthy to the above mentioned practical constraints.

However, machine learning or computational model constructions usually demand large set of (diverse) training instances and each of the training instances with good number of sample points. The classical ImageNet 2012 classification dataset consists 1.28 million training datasets [3]. Popular CIFAR-10 dataset (Canadian Institute For Advanced Research) consists of 50,000 training images [4]. Considering the large number of training instances are used for the validation of a typical learning algorithms, computational model development over time series sensor datasets pose a different kind of challenge.

It is notionally imperative to understand that classification task for practical time series signals require a model that has the ability to minimize the generalization error or minimization of overfitting problem. Hence, our endeavor is to introduce stronger regularization on deep convolutional neural network to ascertain the bias-variance trade-off. While, a deep-layered CNN provides good amount of fitting over training datasets, stronger regularization minimizes that affect in generalization error. Hence, we propose a network construction that is tailor-made for practical analysis of time series sensor signals.

**Table 1 Time series datasets properties**

| Dataset | Number of classes | Number of training instances | Sample points at each training instances |
|---|---|---|---|
| Italy Power Demand | 2 | 67 | 24 |
| Chlorine concentration | 3 | 467 | 166 |
| Cricket X | 12 | 390 | 300 |
| Lightning 2 | 2 | 637 | 60 |
| Coffee | 2 | 28 | 286 |
| Fish | 7 | 175 | 463 |
| Medical Images | 10 | 381 | 99 |

In order to warrant the development of an appropriate classification model for time series sensor signals, we propose SRDCNN: Strongly Regularized Deep Convolution Neural Network architecture to construct a reliable computational model that minimizes the consequences of the practical constraints. We purposefully control the network parameters, mainly the weights of the neural network by both $L^1$ and $L^2$ norm penalties. Traditionally, the conventional deep architecture tends to overfit the training dataset when encountered with smaller number of training instances. We have controlled the overfitting problem by constructing the network with heavily penalized network modeling by incorporating both $L^1$ and $L^2$ norm penalties. The impact of $L^1$ regularization is to ensure sparser weight vector while $L^2$ regularization controls the weight vector values. Thus, SRDCNN attempts to enable a robust learning model against practical time series sensor signals.

This paper is organized as follows. In Section 2, we present the current state of the art methods describing the most relevant algorithms. In Section 3, we briefly illustrate few of the real-life applications where the proposed method would be likely to be applied. Next, in Section 4, we depict the development and deployment architecture of the proposed method. We describe SRDCNN architecture and he network construction method in Section 5. In Section 6, we present results on publicly available UCR datasets. Finally, we conclude in Section 6.

## 2. STATE-OF-THE-ART

Time series classification tasks, especially after the advent of high power computing devices, wearable sensors, deployment of IoT infrastructure and applications become one of the most important as well as challenging research problems. Historically, authors have attempted to use dynamic time warping (DTW) based similarity measures with supervised learning through h k-nearest neighbor (kNN) based classifier [12]. It is described in [1] that DTW is a good benchmark and with some kind of restriction imposed on the amount of warping, DTW with kNN classifer provide reasonable classification performance. Successively, dictionary based methods that learn through transforming the signal to a different domain based on dictionary profile have gained popularity. Symbolic Aggregate Approximation (SAX) -based dictionary classifier, bag of patterns [13] and symbolic Fourier approximation based substructure learning (BOSS) [11] are few of them. Subsequently, more powerful ensemble learning based methods have appeared. Bagnall et al. propose a combination of classifiers in different transformation domains in an ensemble learned way, called Collective of Transformation Ensembles (COTE) Collective of Transformation Ensembles (COTE). Neural network or Multi-Layer Perceptron (MLP)-based methods have also gaining prominence due to the absence of costly feature engineering in shallow learning methods [2].

## 3. USECASES

In this Section, we describe few of the plausible knowledge-driven applications that require computational model based analysis on sensor datasets and impact human quality of life. The main motivation of this research work is to showcase the impact on human-centric applications. In fact, time series sensor signals are integral part of large number of human-centric application. It is to be noted that IoT, powerful learning algorithms, impressive sensing technology and highly powerful computing in low footprint have made different important applications possible in an automated way, minimizing the human effort, cost and maximizing the scalability. We find a gamut of useful human-centric applications to demonstrate the worth of the proposed architecture. For example, remote cardiac health monitoring can be developed by extracting Electrocardiogram (ECG) signals at home from sensors like AliveCor Kardia [5] and analyzing to detect the presence of cardiac abnormality. Automated detection of coffee beans is of immense assistance to minimize the mix of different types of coffee beans. Manually checking large quantity of coffee beans seem to be laborious and error-prone. Another useful application is detect the appliance usage details from domestic electrical energy consumption data. Such analysis and understanding of the electrical consumption can consequently be considered as a way of carbon footprint reduction awareness and campaign. Other than the cited exemplary applications, we feel that large set of human-centric applications require time series sensor analysis. In this paper, we evaluate such types of signals and demonstrate that efficacy of our method. These types of datasets with existing benchmark results are publicly available in UCR time series database [1][2].

## 4. DEVELOPMENT AND DEPLOYMENT ARCHITECTURES

Following in Figure 1 and Figure 2, we depict the development and deployment architectures. The illustrations are conventional. Firstly, the network with pre-defined hyperparameters are trained with training instances consists of a number of labeled time series sensor signals. The trained deep convolution neural network model $M_{SRDCNN}$, with given hyperparameters from the labelled training time series sensor signals is constructed. The model construction is usually performed in the cloud or in high performance computing machines with Graphics Processing Units (GPUs). The trained model $M_{SRDCNN}$ is used for testing or field trial. The trained model can be deployed on normal computing machines including smartphones. The envisioned system development, particularly under IoT-infrastructure, requires data security [17][18] privacy preservation supports like [13][15][19][20] along with secure channel established data transmission. For sensor-initiated communication and autonomous sensor-based applications, wireless sensor networks (WSN) need to deployed [16]. In fact, remote and scalable human-centric applications may typically demand WSN network support.

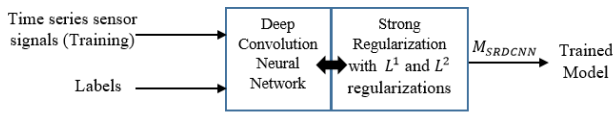

Figure 1. Development architecture for trained model construction.

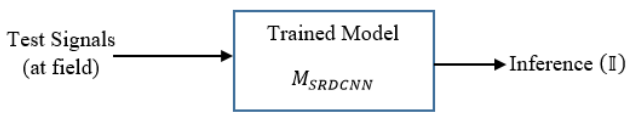

Figure 2. Deployment architecture for on-field testing.

## 5. SRDCNN ARCHITECTURE AND CONSTRUCTION

In this Section, we describe the proposed classification method, related architecture. We first introduce few definitions, which allow us to comprehensively present the relevant methods and approaches.

**Definition 1.** $X = [x_1, x_2, x_3, \ldots, x_T]$ be the univariate time series, where $X \in \mathbb{R}^T$ and $X$ is a time series sensor signal.

**Definition 2.** Dataset $\mathcal{D}$ consists of a collection of pairs $(X^n, L^n)$, where $X^n \in \mathbb{X}$, and $\mathbb{X} = [X^1, X^2, X^3, \ldots X^N]$ consists of a collection of time series signals each of length $T$, $X^n \in \mathbb{R}^T$, $L^n$ is the corresponding class labels, $L^n \in [1, C]$, $C \in \mathbb{Z}$ and $\mathbb{L} = [L^1, L^2, L^3, \ldots L^N]$. Each of the time series is assigned with a class label. Labeled datasets $\mathcal{D}$ is the training dataset.

**Definition 3.** The learning algorithm constructs a function $F: \mathbb{R}^T \to \{1, 2, \ldots, C\}$. The learning algorithm requires the (training) dataset $\mathcal{D}$ and generates trained model $M$. The learning algorithm is further a function of regularization factors $\psi$ and functions $\Upsilon$ along with a collection of necessary hyperparemeters $\Theta$ for constructing the trained model $M$ and trained model is generated as:

$$M \xrightarrow{F: f(\mathcal{D}, \psi, \Upsilon, \Theta)} \hat{L}$$

$\hat{L} \in [1, C]$ is the predicted inference out.

We explore deep convolution neural networks to analyze the time series sensor signals, viz. $X$. Recurrent neural network (RNN) based sequential models or its variants like long short term memory (LSTM) are conventionally used for time series signal analysis [6][7]. RNN or its variants sequentially process the time series to minimize the given loss function $\mathcal{E}$. LSTM like typical RNN backpropagates through time with the capability of minimizing the vanishing gradient problem associated with classical RNN methods. Sequential processing is appropriate where sequential information is closely preserved like in text analysis, speech or audio analysis. Consequently, RNN-like networks have shown substantial success in speech based learning or text analysis [8]. However, we apprehend that typical time series signals lack sequential information and consequently, RNN like model is not suitable. Subsequently, we consider convolution networks for time series sensor analytics. In Figure 3 and Figure 4, we illustrate examples of the time series signals from [1][2]. It is noteworthy that in real-life time series sensor signals, sequential information is not only absent, the signal dynamics seem to follow a non-stationary characteristics. Another problem is the smaller length of the training instances with as low as below 30, $T < 30$. Example, 'Coffee' time series from [1][2]. Owing to such constraints in practical applications, we feel that convolution networks that in image recognition tasks, have achieved near human level performance [9], would be a relevant choice to explore.

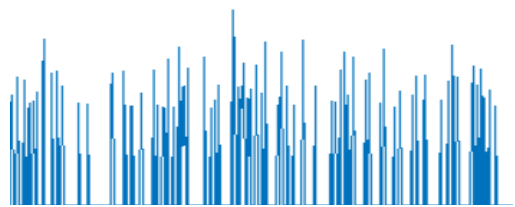

Figure 3. Example of time series sensor signal ('Earthquake' [1][2]: randomly selected training signal instance).

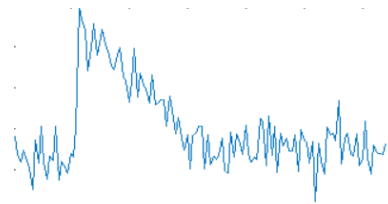

Figure 4. Example of time series sensor signal ('CBF' [1][2]: randomly selected training signal instance).

We demonstrate our method considering 1D time series signals specifically with the intention that CNN type of architecture have shown immensely respectful performance over 2D signals like images, while little study has been made on CNN in 1D signals. In CNN, convolution operation is performed with a filter sliding over $X = [x_1, x_2, x_3, \ldots, x_T]$. CNN is basically a continuous sequence of Convolution Layers, which are interspersed with non-linear functions, called activation functions. In general, the operation is represented as:

$$Z_t = f\left(\omega * X_{\frac{(t-s)}{2}:\frac{(t+s)}{2}} + b\right), \quad \forall t \in T$$

Where, $Z_t$ is the output over the convolution filter of length $s$ at $t$-th time instant for weight $\omega$, bias $b$ with non-linear function $f$ (Rectified Linear Unit-ReLU). There are number of such convolution filters. Convolution filters are followed by a pooling operation, where discriminative filtering is performed. We have used *global average* pooling. In order to reduce error due to bad weight initialization and tackle the problem of covariant shift, batch normalization before each of the activation function (ReLU) layer is introduced. Finally, a fully connected layer is placed. Followed by a final discrimination layer, which is a *softmax* function that predicts the output label $\hat{L}$. The predicted label $\hat{L}$ and actual class label $L$ are compared by a loss function (*cross-entropy*). In the case we consider the loss for the entire training examples $\mathbb{X}$ consisting of $N$ number of training instances, $\mathbb{X} = [X^1, X^2, X^3, ... X^N]$ the cost function $J$ is formed, which is minimized by stochastic gradient descent algorithm. Without regularization in place, an over-complex model is constructed with very high number of parameters (mostly in terms of weight parameters $\omega$). it is unlikely that such model include a closer approximation of the target function or the source data generation function. We control the deep network by introducing regularization as discussed further [14]. Our deep convolution neural network in the training phase minimizes the regularized cost function $\hat{J}$. The regularized cost function is denoted as:

$$\hat{J}(\omega; \mathbb{X}, \mathbb{L}) = J(\omega; \mathbb{X}, \mathbb{L}) + \alpha \Omega(\omega) \quad (1)$$

Where $\hat{J}$ is the regularized cost function to discover the weights $\omega$ from training instance set $\mathbb{X}$, corresponding class labels $\mathbb{L}$ along with penalty function $\Omega$ and penalty factor $\alpha \in [0, \infty]$.

We are interested in parameter norm penalties as expressed above particularly $L^2$ and $L^1$ regularizations.

Through Tikhonov regularization, we incorporate $L^2$ regularization as:

$$\hat{J}(\omega; \mathbb{X}, \mathbb{L}) = J(\omega; \mathbb{X}, \mathbb{L}) + \alpha \frac{\omega^T \omega}{2} \quad (2)$$

The parameter gradient becomes:

$$\nabla_\omega \hat{J}(\omega; \mathbb{X}, \mathbb{L}) = \alpha \omega + \nabla_\omega J(\omega; \mathbb{X}, \mathbb{L}) \quad (3)$$

The weights are updated as:

$$\omega \leftarrow (1 - \epsilon\alpha)\omega + \alpha \nabla_\omega J(\omega; \mathbb{X}, \mathbb{L})$$

Where, $\epsilon$ is the learning rate. In the above expression, we observe the weight decay term $(1 - \epsilon\alpha)$ that regulates the weight vector.

The $L^1$ regularization is defined as:

$$\hat{J}(\omega; \mathbb{X}, \mathbb{L}) = J(\omega; \mathbb{X}, \mathbb{L}) + \alpha \|\omega\|_1 \quad (4)$$

The subsequent parameter gradient is:

$$\nabla_\omega \hat{J}(\omega; \mathbb{X}, \mathbb{L}) = \alpha \text{sign}(\omega) + \nabla_\omega J(\omega; \mathbb{X}, \mathbb{L}) \quad (5)$$

We observe from eq. (3) and eq. (5) that parameter gradients behave differently for $L^2$ and $L^1$ regularizations. It is found that $L^1$ regularization provides sparser weights, while $L^2$ regularization inhibits the weights to gain very high values.

For practical time series signals, we intend a sparser solution such that good number of weights ($\omega$) become zero while the remaining weights are controlled to nominal values. Thus, we propose the following cost function for SRDCNN:

$$\hat{J}_{SRDCNN}(\omega; \mathbb{X}, \mathbb{L}) = J(\omega; \mathbb{X}, \mathbb{L}) + \alpha_1 \|\omega\|_1 + \alpha_2 \frac{\omega^T \omega}{2} \quad (6)$$

Where, $\alpha_1$ is the $L^1$ regularization factor and $\alpha_2$ is the $L^2$ regularization factor.

We depict the HDCNN architecture in Figure 5. Our main novelty is imposing both $L^1$ and $L^2$ regularizations. We apply number of convolution filters over the time series signal. Each of the convolution filter operation is followed by batch normalization and nonlinear activation function ReLU. The feature dimension or the number of neurons at each of the filters is $K_i, i = 1,2,3, ..., l$, where $l$ denotes the number of layers. Batch normalization reduces the covariance shift of the hidden units. Finally, a global average pooling layer is introduced to accommodate a densely connected layer. Global average pooling computes the mean output of each of the previous layer feature map for preparing the model for the final dense or fully connected layer. Softmax function, being close to a categorical probability distribution, so that we can compute categorical cross-entropy based class prediction $\hat{L}$ loss over the true class $L$. However, we compute the regularized cost function $\hat{J}_{SRDCNN}$ over the entire class label space $\mathbb{L}$ as per eq. (6). The weight vector of the proposed SRDCNN is constructed based on eq. (6).

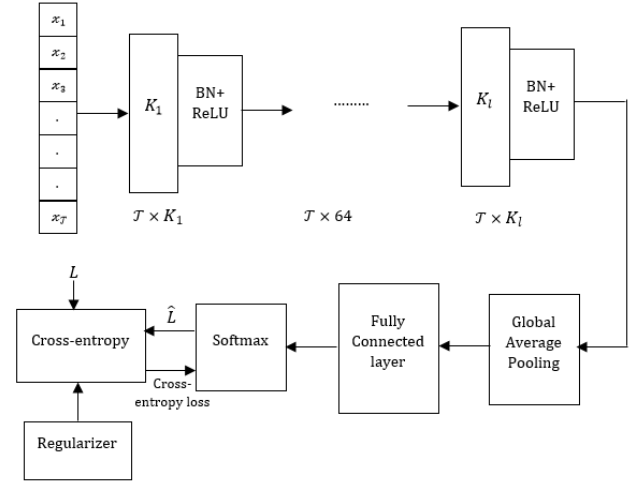

Figure 5. SRDCNN network architecture.

## 6. RESULTS AND ANALYSIS

In order to empirically evaluate the proposed method, we first construct the network as per the described hyperparameters in Table 2. The selected hyperparemeters are typically standard and conventional. The hyperparameters are fixed and the only variable factor is the training instances $\mathbb{X}$ with corresponding label space $\mathbb{L}$. Our contribution in the hyperparameter selection process is that values for $L^1$ regularization factor ($\alpha_1$) and $L^2$ regularization factor ($\alpha_1$). We have considered $\alpha_2 > \alpha_1$. In fact, the value of $\alpha_2$ is twice that of $\alpha_1$. $L^1$ regularization introduces feature selection or sparsity and $L^2$ regularization controls the growth of weight vectors. Intuitively, we require good amount of weight vectors not to be discarded, while the values of the weight vectors should be sufficiently clipped by $L^2$ regularization.

**Table 2 SRDCNN hyperparameter description**

| Parameter | Brief explanation | Value/ Type |
|---|---|---|
| Epoch | Number of times the entire training dataset is checked | 1000 |
| Optimizer | Adaptive learning rate optimization | Adam |
| Batch size | Number of training samples in one pass | min(round($\frac{\mathcal{T}}{10}$), 16) |

| Number of layers | Total number of convolution layers | 5 |
|---|---|---|
| Kernel size | Length of the convolution window | {32,16,8,4,2} |
| Number of filters | Feature map length or number of neurons at each layers. | {32,32,32,32,32} |
| $\alpha_1$ | $L^1$ regularization factor | 0.01 |
| $\alpha_2$ | $L^2$ regularization factor | 0.02 |

In order to empirically understand the efficacy of SRDCNN, we experiment with randomly selected thirteen datasets for UCR time series database [1][2]. These datasets represent heterogeneous types of sensor signals covering diverse set of applications. Trained model of SRDCNN is generated $M_{SRDCNN}$ for each of the datasets from their training instances with the hyperparameters listed in Table 2. In Table 3, we depict the results in terms of accuracy metric over the test instances of the time series data. We compare our obtained results with relevant state-of-the-art algorithms, namely, MLP [2], DTW-R1-1NN [1][2], BOSS [11] and COTE [10]. Currently, BOSS [11] and COTE [10] algorithms are regarded as the benchmark algorithm for time series classification tasks [1]. We further find the number of cases where each of the algorithms perform the best. SRDCNN outperforms the current benchmark algorithms. While SRDCNN achieves the best performance in 6 cases out of 13 cases, next best is COTE[10], which shows 4 best performances over 13 cases.

**Table 3 Performance in terms of accuracy metric of SRDCNN and related stat-of-the-art time series classification algorithms**

| Sensor name | MLP [2] | DTW-R1-1NN[12] | BOSS[11] | COTE [10] | SRDCNN |
|---|---|---|---|---|---|
| CBF | 0.8677 | 0.9933 | 0.9980 | 0.9980 | **0.9989** |
| Coffee | 0.9996 | 0.9889 | 0.9885 | 0.9996 | **1** |
| Large Kitchen Appliances | 0.3565 | 0.7965 | 0.8366 | **0.9000** | 0.8520 |
| Small Kitchen Appliances | 0.3563 | 0.6402 | 0.7502 | 0.7878 | **0.7953** |
| Fish | 0.8647 | 0.7632 | **0.9687** | 0.9622 | 0.9492 |
| CricketX | 0.4994 | 0.7380 | 0.7493 | 0.8149 | **0.8224** |
| Chlorine Concentration | **0.8699** | 0.6455 | 0.6596 | 0.7356 | 0.5433 |
| Lightning2 | 0.7021 | 0.8229 | 0.8100 | 0.7847 | **0.8251** |
| Italy Power Demand | 0.9539 | 0.9226 | 0.8660 | **0.9703** | 0.9546 |
| Haptics | 0.4453 | 0.3900 | 0.4590 | 0.5169 | **0.5247** |
| FordA | 0.7455 | 0.5769 | 0.9195 | **0.9545** | 0.9251 |
| ECG200 | 0.8294 | 0.7987 | **0.8905** | 0.8729 | 0.8791 |
| Medical Images | 0.6657 | 0.7411 | 0.7146 | **0.7850** | 0.6693 |
| Total Count | 1 | 0 | 2 | 4 | 6 |

## 7. CONCLUSION

Sensor signal time series analysis for classification tasks is a critical requirement for different applications that impact the human quality of life. Owing to the heterogeneity of the types of signals and diversity of the applications, a unified model or approach is worthy for practical development and deployment. Our proposed approach ensures a common model to analyze and classify arbitrary time series signals. Currently, we support univariate signals. In future, our endeavor is to develop multi-variate time series signals without disturbing the model architecture. The proposed method of strong regularization based deep network architecture is tailor-made for real-world time series sensor signals in order to address the practical issues and constraints like insignificant number of training instances, smaller number of training samples as well as the development issue of requirement for considerable shorter training time and limited computational resource availability. Our proposed deep architecture further ensures the minimization of generalization error. In fact, the constraints in the training phase often leads the possibility of overfitting problem. Consequently, the proposed method, which is a hyperparameter-controlled deep convolution neural Network, attempts to provide a reliable and practical solution to the real-world challenge encountered in solving the classification tasks for time series sensor signals.

## 8. ACKNOWLEDGEMENT

Leandro Marin is partially supported by Research Project TIN2017-86885-R from the Spanish Ministery of Economy, Industry and Competitivity and Feder (European Union).

Antonio J. Jara is funded by the European Union's Horizon 2020 research and innovation programme under grant agreement No 732679, ACTIVAGE project https://www.activageproject.eu/.

## 9. REFERENCES

[1] Anthony Bagnall, Jason Lines, Aaron Bostrom, James Large and Eamonn Keogh. The Great Time Series Classification Bake Off: a Review and Experimental Evaluation of Recent Algorithmic Advances. Data Mining and Knowledge Discovery, 31(3), 2017.

[2] Anthony Bagnall, Jason Lines, William Vickers and Eamonn Keogh, The UEA & UCR Time Series Classification Repository, www.timeseriesclassification.com

[3] O. Russakovsky, J. Deng, H. Su, J. Krause, S. Satheesh, S. Ma, Z. Huang, A. Karpathy, A. Khosla, M. Bernstein, et al. Imagenet large scale visual recognition challenge. arXiv:1409.0575, 2014.

[4] Alex Krizhevsky, Ilya Sutskever, Geoffrey E Hinton. Imagenet classification with deep convolutional neural networks. Advances in neural information processing systems, pp. 1097-1105, 2012.

[5] https://www.alivecor.com/

[6] Kundan Krishna, Deepali Jain, Sanket V. Mehta, and Sunav Choudhary. 2018. An LSTM Based System for Prediction of Human Activities with Durations. Proc. ACM Interact. Mob. Wearable Ubiquitous Technol. 1, 4, Article 147 (January 2018), 31 pages. DOI: https://doi.org/10.1145/3161201.

[7] Bing Zhang, Jhen-Long Wu, and Pei-Chann Chang. 2018. A multiple time series-based recurrent neural network for short-term load forecasting. Soft Comput. 22, 12 (June 2018), 4099-4112. DOI: https://doi.org/10.1007/s00500-017-2624-5.

[8] Grogory Gelly and Jean-Luc Gauvain. 2018. Optimization of RNN-Based Speech Activity Detection. IEEE/ACM Transactions on Audio, Speech, and Language Processing, vol. 26, no. 3, pp. 646-656, March 2018. doi: 10.1109/TASLP.2017.2769220.

[9] Christian Szegedy et al. Going deeper with convolutions. IEEE Conference on Computer Vision and Pattern Recognition (CVPR), pp. 1-9, 2015. doi: 10.1109/CVPR.2015.7298594.

[10] A. Bagnall, J. Lines, J. Hills, A. Bostrom. 2016. Time-series classification with COTE: The collective of transformation-based ensembles. International Conference on Data Engineering, pp. 1548–1549, 2016.

[11] P. Schafer. 2015.The boss is concerned with time series classification in the presence of noise. Data Mining and Knowledge Discovery, vol. 29, no. 6, pp. 1505–1530, 2015.

[12] E. Keogh and C. A. Ratanamahatana. 2005. Exact indexing of dynamic time warping. Knowledge and information systems, vol. 7, no. 3, pp. 358–386, 2005.

[13] A. Ukil and J. Sen, "Secure multiparty privacy preserving data aggregation by modular arithmetic," 2010 First International Conference On Parallel, Distributed and Grid Computing (PDGC 2010), Solan, 2010, pp. 344-349.

[14] Ian Goodfellow, Yoshua Bengio, and Aaron Courville. 2016. Deep Learning. The MIT Press.

[15] Arijit Ukil, "Privacy Preserving Data Aggregation in Wireless Sensor Networks," 2010 6th International Conference on Wireless and Mobile Communications, Valencia, 2010, pp. 435-440.

[16] Arijit Ukil, "Wireless Sensor Networks," In: Smart Wireless Sensor Networks, Intechopen, 2010.

[17] Arijit Ukil, Soma Bandyopadhyay, Abhijan Bhattacharyya, Arpan Pal, Tulika Bose, "Lightweight security scheme for IoT applications using CoAP," International Journal of Pervasive Computing and Communications, Volume 10, Issue 4, pp. 372-392, 2014.

[18] J. Sen, S. Koilakonda, A. Ukil, "A mechanism for detection of cooperative black hole attack in mobile ad hoc networks," IEEE International Conference on Intelligent Systems, Modelling and Simulation (ISMS), pp. 338-343, 2011.

[19] Arijit Ukil, Jaydip Sen. "Secure multiparty privacy preserving data aggregation by modular arithmetic." IEEE International Conference on Parallel Distributed and Grid Computing (PDGC), (2010): 344-349.

[20] Arijit Ukil, "Privacy preserving data aggregation in wireless sensor networks," 6th International Conference on Wireless and Mobile Communications, pp. 435-440, 2010.